\documentclass[twocolumn,showpacs,preprintnumbers,prl]{revtex4}

\usepackage[dvips]{graphicx}
\usepackage[dvips]{color}

\usepackage{graphicx}
\usepackage{epsfig}

\begin{document}
\title{Analytical Solution of Electron Spin Decoherence Through Hyperfine
Interaction in a Quantum Dot}
\author{Changxue Deng}
\author{Xuedong Hu}
\affiliation{Department of Physics, University at Buffalo, SUNY,
Buffalo, New York 14260-1500}
\date{\today}

\begin{abstract}
We analytically solve the {\it Non-Markovian} single electron spin dynamics
due to hyperfine interaction with surrounding nuclei in a quantum dot.  We use
the equation-of-motion method assisted with a large field expansion, and find
that virtual nuclear spin flip-flops mediated by the electron contribute
significantly to a complete decoherence of transverse electron spin
correlation function.  Our results show that a 90\% nuclear polarization can
enhance the electron spin $T_2$ time by almost two orders of magnitude.  In
the long time limit, the electron spin correlation function has a
non-exponential $1/t^2$ decay in the presence of both polarized and
unpolarized nuclei. 
\end{abstract}
\maketitle


Spins in semiconductor nanostructures are promising qubit candidates for a
solid state quantum computer because of their long decoherence times and
potential scalability \cite{Review}.  
%
%
To demonstrate the feasibility of a spin qubit, electron spin decoherence in
semiconductor quantum dots (QD) has been widely studied both theoretically
and experimentally \cite{Review,LD,Fujisawa,Koppens,Petta}.  The level
discretization in a QD ensures that spin-orbit interaction induced spin
relaxation is quite slow in QDs \cite{Golovach}, which leaves the
environmental nuclear spins, particularly abundant in III-V semiconductors
such as GaAs ($10^4-10^6$ depending on the actual size of the QD), as
the main source of decoherence for the electron spins.  

It has been shown that static thermal polarization of nuclear spins leads to
inhomogeneous broadening of electron spins (which can be corrected using the
spin echo technique \cite{Slichter}) at a time scale of 10 ns
\cite{merk,Koppens,Petta}, and nuclear magnetic dipolar coupling leads to
electron spin spectral diffusion and dephasing at a time scale of 10 $\mu$s
\cite{roger,Witzel}.  For the intervening period of time, the hyperfine
interaction between the electron and nuclear spins can also lead to electron
spin decoherence, which is in general non-Markovian because nuclear dynamics
is slower than the hyperfine dynamics.  

The study of the non-Markovian electron spin dynamics in the presence of
hyperfine interaction is a complicated problem due to its quantum many-body
(1 electron spin and $N$ nuclear spins) nature, and has drawn wide spread
attention recently \cite{khaet, coish, neil,liu}.  Analytically, an exact
solution has been found in the case of a fully polarized nuclear reservoir
\cite{khaet}, while for the rest of the parameter regimes (in terms of
nuclear polarization and external field), perturbative theory \cite{coish} or
effective Hamiltonians \cite{neil,liu} have been used to study the problem. 
Numerically, only small systems with typically less than 20 spins have been
explored because of the extremely large Hilbert space \cite{sch,erl,neil}.  

In this Letter we focus on the problem of the spin decoherence of a single
electron due to hyperfine interaction with the surrounding nuclear spins.
Although at a finite magnetic field the direct electron-nuclear spin flip-flop
is highly unlikely due to the Zeeman energy mismatch, higher-order processes
where electron spins do not flip are possible.  For example, conduction
electron mediated nuclear spin interaction (RKKY) has been studied for a long
time in both metals and semiconductors \cite{old,Slichter}.  Here our focus
is the backaction of the electron-mediated RKKY interaction between nuclear
spins on the {\it single} mediating electron spin.  We start from the exact
electron-nuclear-spin Zeeman and hyperfine Hamiltonian and use the
equation-of-motion approach in the Heisenberg picture.  Helped by a
systematic large field expansion, we solve the full quantum mechanical
problem analytically and reveal the crucial importance of the
electron-mediated nuclear spin flip-flop processes in the decoherence of an
electron spin.

Calculating Green's functions with the equations of motion is an old technique
in solid state physics \cite{mahan}.  What is novel in our current study
is that we use this venerable technique to attack the new problem of spin
decoherence, which is generally studied using quantum master equations for
the density operator \cite{Weiss}.  This traditional approach originating
from quantum optics is more adapted in dealing with weak interactions between
a system and its reservoir.  We demonstrate in this study that a properly
defined correlation function can be used to fully characterize the
decoherence properties of a two-level system, and the equation of motion
approach can be a powerful tool in studying non-Markovian dynamics.

We model the coupled electron-nuclear-spin system by the Hamiltonian
\cite{Slichter}
\begin{equation}
H = \omega_0 S^z + \sum_k A_k I_k^z S^z
  + \frac{1}{2} \sum_k A_k (I_k^+ S^- + I_k^-S^+),
\end{equation}
where $S$ and $I$ represent electron and nuclear spin operators respectively,
$\omega_0$ is the external magnetic field, $A_k$ is the hyperfine coupling
constant with the $k$th nucleus, and $\hbar = 1$.  In this Letter we assume
$I=\frac{1}{2}$ for simplicity, though all calculations can be generalized
for arbitrary $I$.  For a two-dimensional QD with a Gaussian electron wave
function, $A_k$ has the simple form $A_k = A_0 e^{-k/N}$ with $k \in
(0,\infty) $ \cite{coish}.  For convenience we assume $A_0 = 1$ so that time
is measured in the unit of $1/A_0$.  

To describe the decoherence between electron spin $|\uparrow \rangle$ and
$|\downarrow\rangle$ states, we introduce a retarded transverse spin
correlation function
\begin{equation}
G_{\perp}(t) = -i \theta(t)\langle \Psi_0 | S^-(t)S^+(0) |\Psi_0 \rangle \,.
\label{G_t}
\end{equation}
Here $\theta(t)$ is the usual step function, and $|\Psi_0 \rangle$ is the
initial wave function of the system where the electron and nuclear spins are
assumed to be in a product state, with the electron having spin down
initially, {\it i.e.} $|\Psi_0 \rangle = |\Downarrow; I_{k_1}^z, I_{k_2}^z,
\cdots, I_{k_N}^z \rangle$ \cite{note1}.  This spin correlation function
represents the phase fluctuations between electron spin up and down states in
the presence of the nuclear spin reservoir, which can be most clearly seen in
the Schr\"{o}dinger picture
\begin{eqnarray}
G_{\perp}(t) & = & -i \theta(t) \langle \Downarrow; I_{k_1}^z, \cdots,
I_{k_N}^z | e^{iHt/\hbar} S^- e^{-iHt/\hbar} S^+(0) \nonumber \\ 
& & \times | \Downarrow; I_{k_1}^z, \cdots, I_{k_N}^z \rangle \nonumber \\
& = & -i \theta(t) \left\{ \langle \Downarrow; I_{k_1}^z, \cdots, I_{k_N}^z |
e^{iHt/\hbar} \right\} S^- \nonumber \\
& & \times \left\{ e^{-iHt/\hbar} | \Uparrow; I_{k_1}^z, \cdots, I_{k_N}^z
\rangle \right\} \,.
\end{eqnarray}
The term in the first curly bracket represents the evolution of the electron
spin down state in the presence of the hyperfine interaction, while the term
in the second curly bracket represents the evolution of the electron spin up
state in the same environment.  If no electron spin flip occurs, any decay in
the calculated average would be due solely to dephasing between the electron
spin up and down states.  Obviously, electron spin flip will also cause decay
of the correlation function.  Therefore, $G_{\perp}(t)$ contains the complete
decoherence information for the electron spin in consideration.

An iterative equation of motion (EOM) for the spin correlation function
$G_{\perp}(t)$ can be obtained by differentiating $G_{\perp}(t)$ with respect
to time and then perform the Fourier transform.  In general, for two
arbitrary operators $A$ and $B$,
\begin{equation}
\omega \langle\langle A;B\rangle\rangle_\omega = \langle
\Psi_0|A(0)B(0)|\Psi_0 \rangle + \langle \langle [A, H]; B \rangle
\rangle_\omega,
\label{iteration}
\end{equation}
where $\langle\langle A;B\rangle\rangle_\omega$ is the Fourier transform of 
$\langle \Psi_0| A(t)B(0)| \Psi_0\rangle$.  For $G_{\perp}(t)$, we use
$G_{\perp}(\omega)$ to represent its Fourier transform, so that
\begin{equation}
\omega G_{\perp}(\omega) = 1 + \langle \langle [S^-, H]; S^+ \rangle
\rangle_\omega \,.
\label{G_perp(omega)}
\end{equation}
The second terms on the right hand side of Eqs.~(\ref{iteration}) and
(\ref{G_perp(omega)}) involves the calculation of higher-order correlation
functions.  A cut off or decoupling scheme has to be applied to eventually
close the set of EOMs.  

After $G_{\perp}(\omega)$ is obtained, real time dynamics of $G_{\perp}(t)$
can be easily calculated by an inverse Fourier transform using the spectral
function defined as
\begin{equation}
\rho(\omega)=-\text{Im}G_{\perp}(\omega)/\pi.
\label{spec}
\end{equation}
In general there are two types of contribution to the spectral function after
performing analytical continuation ($\omega \rightarrow \omega + i0^+$)
\cite{mahan}: a delta function $Z_p \delta (\omega - \omega_p)$, and a
non-vanishing imaginary part of the self energy resulting from branch cuts,
which results from the integration of continuous poles. 
The delta function leads to a coherent oscillation with a single frequency
$\omega_p$, while the continuous part leads to dephasing in the time
evolution of the spin correlation function $G_{\perp}(t)$.  

We consider the general case of partially polarized and unpolarized nuclear
spin reservoir where both the numbers of spin up and down nuclei are of order
$N$.  The difference in the numbers of the two spin species is characterized
by an effective polarization $P=(N_{\uparrow} - N_{\downarrow})/N$, where
$N_{\uparrow}$ ($N_{\downarrow}$) are the number of nuclear spins in the up
(down) states.  Now the effective magnetic field takes the form $\Omega =
\omega_0 + \sum_kA_k \langle I_k^z \rangle$, where $\langle I_k^z \rangle$
represents time-averaging of $I_k^z(t)$.  
We consider the physically relevant case of large effective fields ($\Omega
\sim N$, requiring that we have either a reasonably large external field, or
a nuclear reservoir with finite polarization), and focus on the spectral
broadening near $\omega = \Omega$, which leads to dephasing of transverse
electron spin magnetization.  

Previous studies \cite{khaet,coish} indicate that when only the direct
electron-nuclei spin flip-flop is considered, the decay amplitude of 
the electron spin correlation function is of the order $O(1/N)$, and the
correlation function has almost undamped oscillations. Clearly, such direct
processes are energetically unfavorable in high effective magnetic fields.
However, if the higher-order virtual process (electron mediated nuclear spin
flip-flop) is included, we expect that nuclear field fluctuation will give
rise to complete decoherence in the electron spin.  In other words, the delta
function (indicating no damping) in the spectral function would be broadened
(decoherence) after the virtual processes are included.  The spectral weight
in the low energy region where $\omega \sim O(1)$ has been found to be
negligible \cite{deng}.
 
%

In the following calculation we treat the nuclear field $\sum_k A_k I_k^z$
within the adiabatic approximation, which is physical since $S^-(t)$
has an oscillation frequency $\Omega \sim N$ while the nuclear field varies
in a much longer time scale, so that in the Fourier transform of $\langle
\Psi_0 |\sum_k A_k I_k^z(t) S^-(t) S^\dagger (0) | \Psi_0 \rangle$ we can
simply replace $\sum_k A_k I_k^z (t)$ by $\sum_k A_k \langle I_k^z \rangle$. 
$G_{\perp}(\omega)$ is related to the higher-order correlation function
$\langle \langle I_k^- I_{k{'}}^{\dagger} S^-; S^\dagger \rangle
\rangle_\omega$ through the following equation
\begin{equation}
\left(\omega - \Omega - \frac{N}{8\Omega}\right) G_{\perp}(\omega) = 1 +
\frac{1}{2\Omega} \sum_{k\neq k^{'}} A_k A_{k^{'}} \langle \langle
I_{k}^{-} I_{k^{'}}^{\dagger} S^-; S^\dagger \rangle\rangle _{\omega} \,.
\end{equation}
Here $\langle \langle I_k^- I_{k{'}}^{\dagger} S^-; S^\dagger \rangle
\rangle_\omega$ represents nuclear spin $k$ and $k^{'}$ flip-flopping with
each other while electron spin returning to its initial state ($\downarrow$). 

\begin{figure}[t]
\begin{center}
\epsfig{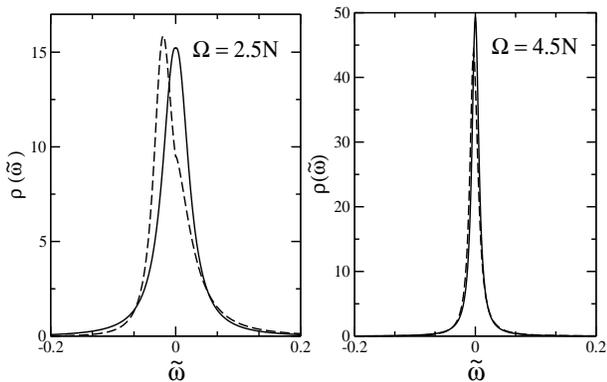}
\caption{
Spectral function $\rho(\tilde{\omega})$ (where $\tilde{\omega} = \omega -
\Omega - N/8\Omega$) with $P=0$ for $\Omega=2.5N$ (left panel) and
$\Omega=4.5N$ (right panel).  In both panels the solid lines represent the
spectral functions calculated by considering only self-energy $\Sigma_1
(\tilde{\omega})$.  The dashed lines show the total contributions from 
$\Sigma_1(\tilde{\omega})$ and $\Sigma_2(\tilde{\omega})$. The original 
delta functions $\delta(\tilde{\omega})$ in the spectral densities are
broadened due to the effects of the flip-flopping of nuclear spins.
}
\label{fig1}
\end{center}
\end{figure}

Calculating this higher-order correlation function requires a cut-off to
terminate the iteration.  The structure of the iterative equations reveals
that a natural cut-off does exist. 
The correlation functions with one
flip-flopped nuclear spin pair has two contributions to the self-energy, one
of them proportional to $N^2/(4\Omega)^2$, the other $N^3/(4\Omega)^3$.  For
two pairs of flip-flopped nuclear spins the two contributions are
proportional to $N^4/(4\Omega)^4$ and $N^5/(4\Omega)^5$ respectively.  This
geometrical series converges quite fast when $\Omega > N$ ({\it large field
expansion}). Physically, the expansion parameter $N/\Omega$ appears because
the intermediate high energy state where the electron spin is flipped requires
energy $\Omega$, and $N$ comes from the summation over all nuclear spins. 
In the limit of $\Omega \gg N$, only the first-order term of
the self-energy contributes significantly.  Neglecting $N^4/\Omega^4$ and
higher-order terms, the final expression of the spin correlation function is
\begin{equation}
G_{\perp}(\omega) = \frac{1}
{\tilde{\omega} -\frac{(P^2-1)N^2}{16\Omega^2}\Sigma_1(\tilde{\omega})
-\frac{(1-P^2)N^3}{32\Omega^3}\Sigma_2(\tilde{\omega})},
\label{G_p}
\end{equation}
with the lowest-order self-energy taking the form
\begin{eqnarray}
\Sigma_1(\tilde{\omega}) &=& \frac{2}{3}\left[ 
\tilde{\omega}(4\tilde{\omega}^2 - 3)\text{log}
\left| 1-\frac{1}{4\tilde{\omega}^2}\right|
+ \tilde{\omega} 
+\text{log}\left|\frac{2\tilde{\omega} - 1}{2\tilde{\omega}+1}\right|
\right] \nonumber \\
&+& i\frac{2\pi}{3} \left[4|\tilde{\omega}|^3 - 3|\tilde{\omega}| +1 \right],
\end{eqnarray}
for $|\tilde{\omega}|<1/2$.  Here $\tilde{\omega} = \omega - \Omega -
N/8\Omega$.  The exact form of $\Sigma_2(\tilde{\omega})$ is also found
\cite{deng}.  Both self-energy terms have branch cuts or non-vanishing
imaginary parts when $|\tilde{\omega}| < 1/2$, leading to dephasing when
calculating $G_\perp (t)$.  Another significant feature of $G_{\perp}
(\omega)$ is that it does not have a $\delta$-function component anymore,
indicating that the decoherence of $G_\perp (t)$ will be complete.  In
addition, Eq.~(\ref{G_p}) indicates that the amplitude of $G_{\perp}(\omega)$
is $\sim O(1)$, in contrast to the fully polarized case, where $G_{\perp}
(\omega) \sim O(1/N)$.

\begin{figure}[t]
\begin{center}
\epsfig{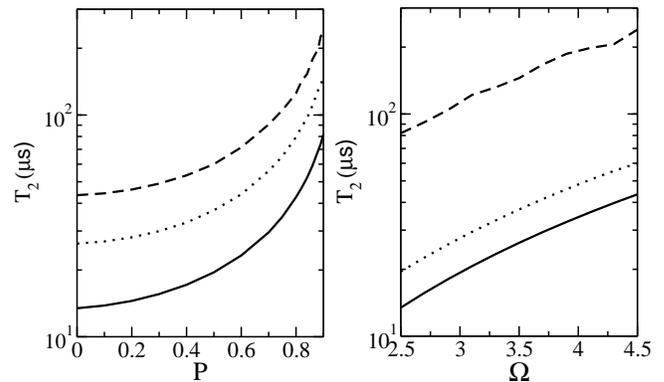}
\caption{ 
$T_2$ determined from the half width of the peak of spectral functions for
different nuclear polarizations ($P$) and effective fields ($\Omega$). In all
cases the spectral functions are calculated using both $\Sigma_1
(\tilde{\omega})$ and $\Sigma_2(\tilde{\omega})$. We have chose $N=10^5$
and $A=\sum A_k=92 ~\mu\text{eV}$.
}
\label{fig2}
\end{center}
\end{figure}

Figure~\ref{fig1} shows the calculated electron spin spectral functions for
different effective fields.  We compare the results of including only
$\Sigma_1(\tilde{\omega})$ (solid lines) and those with both $\Sigma_1
(\tilde{\omega})$ and $\Sigma_2 (\tilde{\omega})$ (dashed lines) for $\Omega
= 2.5 N$ and $\Omega = 4.5 N$.  The two panels clearly show the validity of
the large field expansion for $\Omega \ge 2.5N$.  For smaller $\Omega$ more
higher-order terms need to be included to attain convergence.  Indeed, even
if $\Omega < N$ there is no divergence in our theory, since there could be at
most $N_{\downarrow}~ (N_{\downarrow} < N_{\uparrow})$ flip-flopped nuclear
pairs in the system, so that there is an upper limit to the number of EOMs
and terms in self-energy.  The right panel of Fig.~\ref{fig1} (for $\Omega =
4.5N$) shows that the contribution of $\Sigma_2(\tilde{\omega})$ is now
completely negligible.  
%
%
Using hyperfine coupling constant of bulk GaAs \cite{paget}, we estimate that
$\Omega = 2.5N$ corresponds to a magnetic field of 5 Tesla. 
Figure~\ref{fig1} also explicitly shows that the original delta function in
the spectral function is now broadened after taking into account the
electron-mediated flip-flop of nuclear spins.  According to Eq.~(\ref{G_p}),
in the limit $\Omega \gg N$ or $P=1$, both self-energy terms go to zero, so
that the delta function form of $\delta(\tilde{\omega})$ of the spectral
functions would have been recovered, and there would have been no decoherence
effect.

The decoherence time $T_2$ for the electron spin can be determined from the
half-width ($\Delta \tilde{\omega}$) of the spectral peak ($T_2 = 1/\Delta
\tilde{\omega}$).  Figure \ref{fig2} shows $T_2$ as functions of the nuclear
spin polarization $P$ and the effective magnetic field $\Omega$.  It is clear
that $T_2$ only increases slowly with the external magnetic field, but is much
more sensitive to the nuclear polarization.  If the nuclear polarization $P$
is raised to 0.9 from 0, $T_2$ increases by almost two orders of magnitude. 
Physically this is quite reasonable, as increasing polarization would reduce
the phase space for nuclear spin flip-flops, while increasing external field
slowly reduces the cross-section of these processes.
%

\begin{figure}[t]
\begin{center}
\epsfig{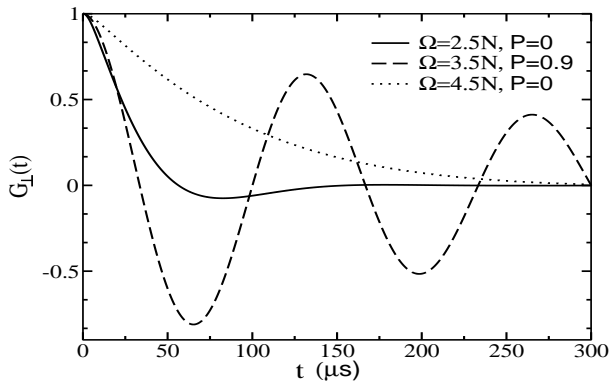}
\caption{ 
Time evolution of the envelope of ${\rm Re} \{G_{\perp}(t)\}$ for various
regimes of $T_2$.  $G_{\perp}(t)$ is obtained from the spectral function by
inverse Fourier transform.  It should be noted that the real evolution is
modulated by the fast oscillation term $e^{i\Omega t}$.  We have assumed 
$N=10^5$ and $A = \sum A_k = 92~\mu \text{eV}$ in these calculations.
}
\label{fig3}
\end{center}
\end{figure}

The real-time dynamics of $G_{\perp}(t)$ is obtained with the inverse
Fourier transform $G_{\perp}(t) = -i\theta(t)\int \rho(\omega) e^{-i\omega t}
d\omega$ using the spectral function calculated with both $\Sigma_1
(\tilde{\omega})$ and $\Sigma_2(\tilde{\omega})$.  Figure~\ref{fig3} plots
the time evolution of the envelope of ${\rm Re}\{G_{\perp}(t)\}$ for
three different parameter regimes.  The solid line represents the case of
fast decay with $\Omega=2.5 N$ and no polarization.  The dotted line shows
that increasing the magnetic field can increase the coherence time
moderately.  If the nuclei in the QD are polarized to 90\%, the amplitude of
the fast oscillation in electron spin (with frequency $\Omega$) could be
maintained for a much longer time as indicated by the dashed line.  Notice
that here the amplitude of $G_\perp (t)$ does decrease as quickly as in the
previous two cases initially, but electron spin quantum coherence is only
partially lost so that clear revival phenomenon is visible after even several
$\mu$s in Fig.~\ref{fig3}.
%

The long-time asymptotic behavior of $G_{\perp}(t)$ can also be extracted from
$\rho(\omega)$, which is nonzero only when $|\tilde{\omega}| < 1/2$. 
Calculating the inverse Fourier transform at $t \gg 1$, we find 
\begin{equation}
G_{\perp}(t) \propto \frac{144\Omega^2}{\pi^2(1-P^2)N^2} \frac{1}{t^2}.
\end{equation}
The $1/t^2$ power-law decay here can be compared to the $1/t$ power-law decay
found in Ref.~\cite{khaet, coish} for large magnetic fields, where
electron-mediated nuclear spin flip-flops are not taken into account, and the
exponential decay found in Ref.~\cite{liu}, where an effective Hamiltonian
for the nuclear spin flip-flop is considered.

In summary, we have presented a detailed analytical study of transverse
electron spin decoherence using large field expansion.  We find that
electron-mediated nuclear spin flip-flops contribute significantly to
electron spin dephasing by generating fluctuations in the Overhauser field
(the nuclear field) for the electron spin.  We find that 80-90\% nuclear
polarization can enhance the electron spin $T_2$ time by two orders of
magnitude into the $\mu$s time scale in a 5 T external field.  We also show
that the long time asymptotic behavior of the spin decoherence is $1/t^2$.

We acknowledge financial support by NSA, LPS, and ARO.


\vspace*{-0.3in}
\begin{thebibliography}{99}

\bibitem{Review} X. Hu and S. Das Sarma, Phys. Stat. Sol. (b) {\bf 238},
360 (2003); X. Hu, cond-mat/0411012, Springer Lect. Notes Phys. {\bf 689},
83-114 (2006); S. Das Sarma {\it et al.}, Solid State Commun. {\bf 133}, 737
(2004).

\bibitem{LD} D. Loss and D.P. DiVincenzo, Phys. Rev A {\bf 57}, 120 (1998).


\bibitem{Fujisawa} T. Fujisawa {\it et al.}, Nature {\bf 419}, 278 (2002).

\bibitem{Koppens} F.H.L. Koppens {\it et al.}, Science {\bf 309}, 1346 (2005).


\bibitem{Petta} J.R. Petta {\it et al.}, Science {\bf 309}, 2180 (2005).

\bibitem{Golovach} V.N. Golovach, A. Khaetskii, and D. Loss, Phys. Rev. Lett.
{\bf 93}, 016601 (2004).

\bibitem{Slichter} C.P. Slichter, {\em Nuclear Magnetic Resonance} 
(Spinger-Verlag, Berlin, 1996).

\bibitem{merk} I.A. Merkulov, A.L. Efros, and M. Rosen, Phys. Rev. B {\bf 65},
205309 (2002).

\bibitem{roger} R. de Sousa and S. Das Sarma, Phys. Rev. B {\bf 68}, 115322
(2003).

\bibitem{Witzel} W.M. Witzel, R. de Sousa, and S. Das Sarma, Phys. Rev. B.
{\bf 72}, 161306(R) (2005).


\bibitem{khaet} A.V. Khaetskii, D. Loss, and L. Glazman, Phys. Rev. Lett. {\bf
88}, 186802 (2002); Phys. Rev. B {\bf 67}, 195329 (2003).

\bibitem{coish} W.A. Coish and D. Loss, Phy. Rev. B {\bf 70}, 195340 (2004).

\bibitem{neil} N. Shenvi, R. de Sousa, and K.B. Whaley, Phys. Rev B {\bf 71},
224411 (2005).

\bibitem{liu} W. Yao, R.B. Liu, and L.J. Sham, cond-mat/0508441.

\bibitem{sch} J. Schliemann, A.V. Khaetskii, and D. Loss, Phys. Rev B
{\bf 66}, 245303 (2002).

\bibitem{erl} S. I. Erlingsson and Y. V. Nazarov, Phys. Rev. B {\bf 70},
205327 (2004).

\bibitem{old} N. Bloembergen and T.J. Rowland, Phys. Rev. {\bf 97}, 1679
(1955).


\bibitem{mahan} G.D. Mahan, {\it Many-Particle Physics} (Plenum Press, New
York, 1990).

\bibitem{Weiss} U. Weiss, {\it Quantum Dissipative Systems} (World Scientific,
Singapore, 2001).

\bibitem{note1} We assume that the nuclear spins are initially in a
product state.  For mixed initial nuclear states, the result of this paper can
be directly used in the sum over all the product states in the ensemble.



\bibitem{deng} C. Deng and X. Hu, in preparation (2005).



\bibitem{paget} D. Paget, G. Lampel, and B. Sapoval, Phys. Rev. B {\bf 15},
5780 (1977).


\end{thebibliography}
\end{document}